# Cluster glass behavior of frustrated birnessites $A_x$MnO$_2$·$y$H$_2$O ($A$ = Na, K)

L. D. Kulish,* R. Scholtens, G. R. Blake

*Zernike Institute for Advanced Materials, University of Groningen, Nijenborgh 4, 9747 AG, Groningen, the Netherlands*

We report on the synthesis and magnetic properties of frustrated Na$_{0.22}$MnO$_2$·0.39H$_2$O and K$_{0.6}$MnO$_2$·0.48H$_2$O with the birnessite structure. The structure, static and dynamic magnetic properties of the compounds are investigated in detail. A combination of DC and AC magnetic susceptibility measurements and magnetization decay measurements reveal cluster glass behavior below the freezing temperature of 4 K for Na-birnessite and 6 K for K-birnessite. The frequency dependence of the freezing temperature is analyzed on the basis of dynamic scaling laws including the critical slowing down formula and the Vogel-Fulcher law, which further confirm cluster glass formation in both compounds.

## I. INTRODUCTION

Magnetically frustrated compounds are promising for the emergence of various exotic magnetic states. For example, they can exhibit spin liquid and spin ice behavior, act as valence-bond solids, or exhibit an array of helical and cycloidal spirals or even a variety of periodic states with non-trivial topologies composed of skyrmions and antiskyrmions.[1,2,3] Magnetic frustration often results from competition between ferromagnetic (FM) and antiferromagnetic (AFM) exchange interactions in crystal lattices based on triangles or tetrahedra that share corners, edges or faces.[4,5]

One group of interesting compounds from this point of view is the alkali manganites $A_x$MnO$_2$·$y$H$_2$O ($A$ = Na, K) with the birnessite structure (hereafter referred to as Na/K-bir). These antiferromagnetic compounds have frustrated 2D triangular planes of magnetic Mn$^{3+/4+}$ ions, which form edge-shared MnO$_6$ octahedral structural units. The planes are separated from each other by gaps containing non-magnetic $A^+$ cations and H$_2$O molecules.[6] This type of structure, with weakly coupled planes of spins pointing typically in the out-of-plane direction, allows tuning of the interlayer distance, the ratio of Mn$^{3+}$/Mn$^{4+}$ and thus the magnetic exchange and anisotropy along the stacking direction.

The exact crystal structure of birnessites remains unclear in terms of the placement of the interlayer species and the presence of Mn vacancies.[7,8] Due to high ionic mobility, it is difficult to determine whether the alkali cations and water molecules occupy the same or different positions in the interlayer space. At the same time, this high mobility of the interlayer cations has led to the wide use of birnessite compounds in the field of battery storage as capacitors showing high cycling capability.[9,10,11] Furthermore, birnessite compounds have been demonstrated as molecular sieves for purposes such as water purification.[12] Concerning vacancies in the manganese layers, it has been suggested that their existence probably depends on synthesis conditions.[8]

Relatively few studies have been reported on the magnetic properties of birnessite compounds. Birnessite-like MnO$_2$ nanowalls exhibited antiferromagnetic (AFM) behavior with an ordering transition at 9.2 K and a bifurcation of the zero-field-cooled and field-cooled DC susceptibilities.[13] No information is available on K-containing birnessites and only one study has been performed on water-containing Na-birnessite structures, namely for Na$_{0.36}$MnO$_2$·0.2H$_2$O.[14] This might be due to the difficulty in synthesizing phase-pure birnessite samples; powder X-ray diffraction (XRD) on the samples studied in Ref. 14 revealed impurities of α- and β-NaMnO$_2$, and possibly Mn$_3$O$_4$, all of which are magnetic phases. It was shown that a spin glass state is present below 29 K and it was speculated that there is a random static distribution of Mn$^{3+}$/Mn$^{4+}$ cations in the matrix of the birnessite compound However, the influence of the impurity phases on the magnetic data in this study was not discussed.

For a better understanding of the nature of birnessite compounds, it should also be mentioned that a closely related type of layered manganese oxides is known – the α and β-NaMnO$_2$ phases without crystal water in the interlayer space. These compounds have been well investigated in terms of their electrical and magnetic properties. Because they contain Mn$^{3+}$ – a strongly Jahn-Teller active cation, the MnO$_6$ octahedra are distorted along the [−1, 0, 1] crystallographic direction. The α-NaMnO$_2$[15,16] phase adopts the monoclinic C2/m space group and consists of flat sheets of edge-shared MnO$_6$ octahedra separated by Na$^+$ ions, similar to Na-bir. Another similarity is the possibility of Na vacancies in the structure which would also lead to mixed-valent manganese ions, but in the birnessite structure the Na/Mn molar ratio is less than 0.7 whereas for α-NaMnO$_2$, the ratio is > 0.7. The β-NaMnO$_2$[17] phase differs from the α-phase in that the MnO$_6$ octahedra form zig-zag sheets separated by Na atoms; the space group is orthorhombic Pmmn.[18]

The α-phase polymorph Na$_{0.9}$MnO$_2$ has unfrustrated nearest-neighbor manganese atoms with a dominant AFM exchange interaction along the short $b$ axis and frustrated next-nearest neighbor AFM exchange due to four equivalent exchange pathways. This frustrated interchain coupling leads to quasi-1D magnetic interactions[19] in this compound at temperatures as high as 200 K. Below the Néel temperature of 45 K, the triangular lattice hosts the coexistence of a short-range incommensurately modulated AFM state with a short-range commensurate AFM state which becomes dominant below 22 K. The presence of a small amount of Mn$^{4+}$ cations is manifested as a random, static magnetic impurity within the MnO$_6$ planes.[15]

In the case of β-NaMnO$_2$, it was shown that a spontaneous long-range collinear AFM order appears below 200 K. Moreover, a transition to a spatially modulated proper screw magnetic state was found at 95 K. Between these two transitions a magnetically inhomogeneous state exists, and a spin gap (~ 5 meV) opens in the low-temperature state.[20]

Summarizing current knowledge of these systems, a precise description of the magnetic properties of birnessite compounds remains obscure, specifically, the influence of the mixed oxidation state of the manganese ions, the distribution of Mn$^{3+/4+}$ in the matrix, as well as how the alkali cation deficiency and the number of water molecules can be controlled to tune the magnetic properties. Here we take one of the first steps to explore the structure and magnetic behavior of birnessite compounds. We utilize sol-gel synthesis to obtain pure samples of the birnessite compounds $A_x$MnO$_2$·yH$_2$O ($A$ = Na, K). Powder XRD shows that the Mn$^{3+}$/Mn$^{4+}$ cations are randomly distributed within the MnO$_6$ layers. Na-bir has a monoclinic structure with space group C2/m, whereas K-bir is triclinic with space group C$\bar{1}$. We use DC and AC magnetic susceptibility studies together with magnetization decay measurements to show that both systems have a glassy nature below their freezing temperatures. The frequency dispersion of the temperature-dependent AC susceptibility can be described by dynamic scaling theory and the Vogel-Fulcher law, which identify these systems as cluster glasses.

## II. EXPERIMENTAL DETAILS

Bulk Na/K-bir samples were prepared by a sol-gel process earlier described by Ching *et al.*[21] First, TEAMnO$_4$ was synthesized by adding tetraethyl-ammonium bromide (TEABr) to KMnO$_4$ (1:1 molar ratio) in water.[22] The solution was stirred for 24 hours after which the precipitate of TEAMnO$_4$ was collected and dried under vacuum at room temperature to avoid thermal degradation. The typical yield of the reaction was 40 % due to the partial water solubility of the salt. Next, 0.4 g (1.1 mmol) of TEAMnO$_4$ was added to 6 ml of a 0.9 M Na/K acetate in methanol solution. The solution slowly turned from purple to red and eventually to brown after ~ 1.5 hours at room temperature. The color change results from the reduction of the Mn$^{7+}$ ion to Mn$^{4+}$/Mn$^{3+}$, which is paired with the oxidation of methanol. During this process the sol forms a gel. Excess methanol was then decanted and the gel was dried at 110 °C for 24 h forming a xerogel. The xerogel was calcined at 450 °C for 2 h to yield a dark-brown powder. After the powder was washed with water (3 times for K-bir and 20 times for Na-bir due to its less hygroscopic nature), it was dried at 50 °C.

Various characterization methods were used to probe the formation of Na/K-bir and to give information on the structural and magnetic properties. The phase purity and the crystal structure of the products were determined by X-ray diffraction. Powder XRD was carried out on a Bruker D8 Advance diffractometer operating with Cu Kα radiation. Measurements were performed in the 2θ range 10-80°. The XRD data were fitted by Rietveld refinement using the GSAS software. To establish the particle shape and the size of birnessite compounds, scanning electron microscopy (SEM, Fei Helios G4 CX DualBeam) was performed. The dehydration processes on heating were investigated by means of simultaneous thermogravimetric analysis (TG) and differential scanning calorimetry (DSC) on a TG 2960 SDT instrument using an argon flow of 100 mL/min; the heating rate was 10 °C/min over the temperature range 30 °C to 500 °C. Magnetic measurements were performed on a Quantum Design MPMS SQUID-based magnetometer. Magnetic susceptibility scans were performed on warming over the range 2-200 K, and magnetization versus applied field curves were obtained between -7 T and 7 T at 2.5 K for Na-bir and 5 K for K-bir. AC susceptibility measurements were performed using a 3.8 Oe oscillating field superimposed on a 200 Oe DC field. Magnetization decay experiments were performed by applying a 1 T field, cooling the sample to 25 K at 1 K/min, holding for 10 min, and then cooled at 1 K/min to 2.5 K for Na-bir and 5 K for K-bir (below the glass freezing temperature). After 1 min, the field was removed and the magnetization was measured as a function of time.

## III. RESULTS AND DISCUSSION

### A. Structural characterization

The chemical compositions of the compounds were determined by a combination of simultaneous thermal analysis (TG and DSC) and a back-titration method (details are given in Appendix A). The calculated amount of water and average manganese oxidation state (+3.78 for Na-bir and +3.37 for K-bir assuming no oxygen or manganese vacancies) yield stoichiometries of Na$_{0.22}$MnO$_2$·0.39H$_2$O and K$_{0.6}$MnO$_2$·0.48H$_2$O.

Structural analysis of our Na/K-bir samples using powder XRD shows single-phase products in both cases. The 001 peaks at 2θ ≈ 2.5° are consistent with the ~7 Å inter-layer spacing expected for the birnessite compounds.[14,23] The space group is monoclinic C2/m for Na-bir and triclinic C$\bar{1}$ for K-bir. The refined lattice parameters are listed in Table I, and details of the XRD analysis are given in Appendix B. In both structures manganese cations occupy a single crystallographic position, which implies randomly distributed Mn$^{3+}$/Mn$^{4+}$ cations within the MnO$_6$ layers (Fig. 1).

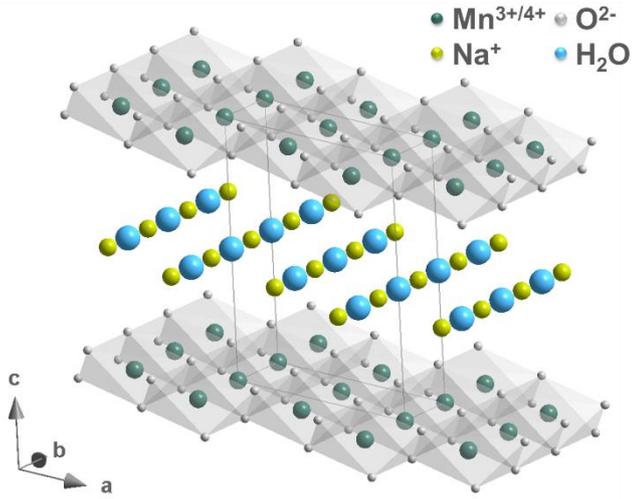

FIG. 1. Schematic representation of the crystal structure of $Na_{0.22}MnO_2 \cdot 0.39H_2O$. The manganese, oxygen and sodium atoms are represented by dark green, grey, and green-yellow spheres respectively; the $H_2O$ molecules are represented by the blue spheres. One unit cell is indicated by the dark grey lines.

TABLE I. Compositions and structural parameters of Na/K-bir samples.

| Compound | Space Group | a | b | c | α | β | γ | Volume |
|---|---|---|---|---|---|---|---|---|
| $Na_{0.22}MnO_2 \cdot 0.39H_2O$ | C2/m | 4.923(1) | 2.857(1) | 7.277(2) | 90 | 102.01(2) | 90 | 100.09 (4) |
| $K_{0.6}MnO_2 \cdot 0.48H_2O$ | C$\bar{1}$ | 5.006(2) | 2.878(1) | 7.281(2) | 88.84(3) | 101.21(3) | 89.33 (4) | 102.89 (5) |

The SEM images in Fig. 2 show that the sol-gel process led to the formation of irregular agglomerates of flat, round particles with different sizes. In the case of K-bir the particle sizes range from 25 to 100 nm, where the Na-bir particles are bigger with sizes from 45 to more than 200 nm.

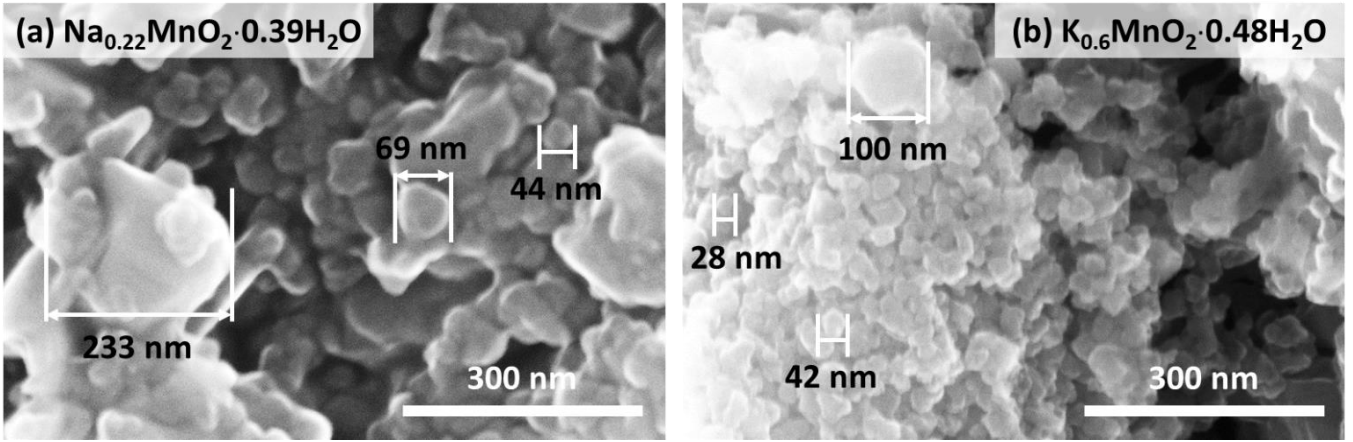

FIG. 2. SEM images of (a) Na-birnessite, (b) K-birnessite.

### B. DC magnetic susceptibility

The magnetic properties of Na/K-bir were initially investigated by performing DC magnetization measurements. Zero-field-cooled (ZFC) and field-cooled (FC) measurements were performed in an applied magnetic field of 500 Oe on warming over the temperature range 2-200 K. A dependence on the thermal-magnetic history of the samples was observed, namely a bifurcation of the FC and ZFC curves below a characteristic temperature $T_{irr}$ (Fig. 3(a)). Such splitting can arise from a variety of phenomena such as spin glass/cluster glass (SG/CG), spin liquid, superparamagnetic, disordered antiferromagnetic and spin spiral states.[24] The FC magnetization shows a continuous increase upon lowering the temperature. Simultaneously, the ZFC curve displays a well-defined peak at a temperature $T_g = 3.5$ K for Na-bir and 6.5 K for K-bir (Table II). The position of $T_g$ is slightly below $T_{irr}$ for both compounds. This behavior can be a manifestation of magnetic clusters, because for a canonical SG it generally holds that $T_{irr} \lesssim T_g$.[25,26]

TABLE II. Summary of magnetic parameters of Na/K-bir.

| Compound | $T_g$ (K) | $T_{irr}$ (K) | $\theta_{CW}$ (K) | $\mu_{eff}$ ($\mu_B$) | $\delta$ | $M_0$ (emu/mol) | $\alpha$ |
|---|---|---|---|---|---|---|---|
| $Na_{0.22}MnO_2 \cdot 0.39H_2O$ | 3.5 | 7.5 | -85 | 3.76 | 0.015 | $1.85 \times 10^{-5}$ | $9.5 \times 10^{-2}$ |
| $K_{0.6}MnO_2 \cdot 0.48H_2O$ | 6.5 | 9.5 | -63 | 3.40 | 0.027 | $4.98 \times 10^{-5}$ | $1.7 \times 10^{-2}$ |

The inverse susceptibility of the samples (Fig. 3(b)) is linear above 50 K; at lower temperature it starts to deviate, most likely due to short-range interactions. The extracted negative Curie-Weiss temperature $\theta_{CW}$ (Table II) suggests that the interactions are antiferromagnetic in both cases. The effective moment $\mu_{eff}$ is 3.76 $\mu_B$ for Na-bir and 3.40 $\mu_B$ for K-bir. The theoretical spin-only $\mu_{eff}$ for $Mn^{4+}$ is 3.88 $\mu_B$; $\mu_{eff}$ is 2.83 $\mu_B$ for the $Mn^{3+}$ low-spin state and 4.9 $\mu_B$ for the $Mn^{3+}$ high-spin state.[24] Our measured values imply that the Mn cations adopt a mixed-valent configuration of $Mn^{4+}$ and low-spin $Mn^{3+}$. Since the presence of $Mn^{4+}$ can suppress the Jahn-Teller distortion associated with $Mn^{3+}$ under octahedral crystal fields, the low-spin state can become favored.[27] This configuration yields average oxidation states calculated from $\mu_{eff}$ of $Mn^{3.7+}$ for Na-bir and $Mn^{3.35+}$ for K-bir, which are in good agreement with the corresponding average oxidation states obtained by the back-titration method of $Mn^{+3.78}$ and $Mn^{+3.37}$ respectively. We note that a previous study of $Na_{0.36}MnO_2 \cdot 0.2H_2O$ also identified a low-spin $Mn^{3+}$ state.[14]

The presence of magnetic frustration can be inferred from the frustration parameter $f$, which is the ratio of $|\theta_{CW}|$ to the magnetic ordering temperature. However, this parameter is only valid if a long-range-ordered state is reached at some temperature, which does not seem to be the case above 2 K (the limit of our measurements) for Na/K-bir. The frustration parameter can be approximated as $f = |\theta_{CW}|/T_g$.[5] For both our compounds, $|\theta_{CW}|$ is 10 or more times greater than $T_g$, implying a high level of frustration (Table II).

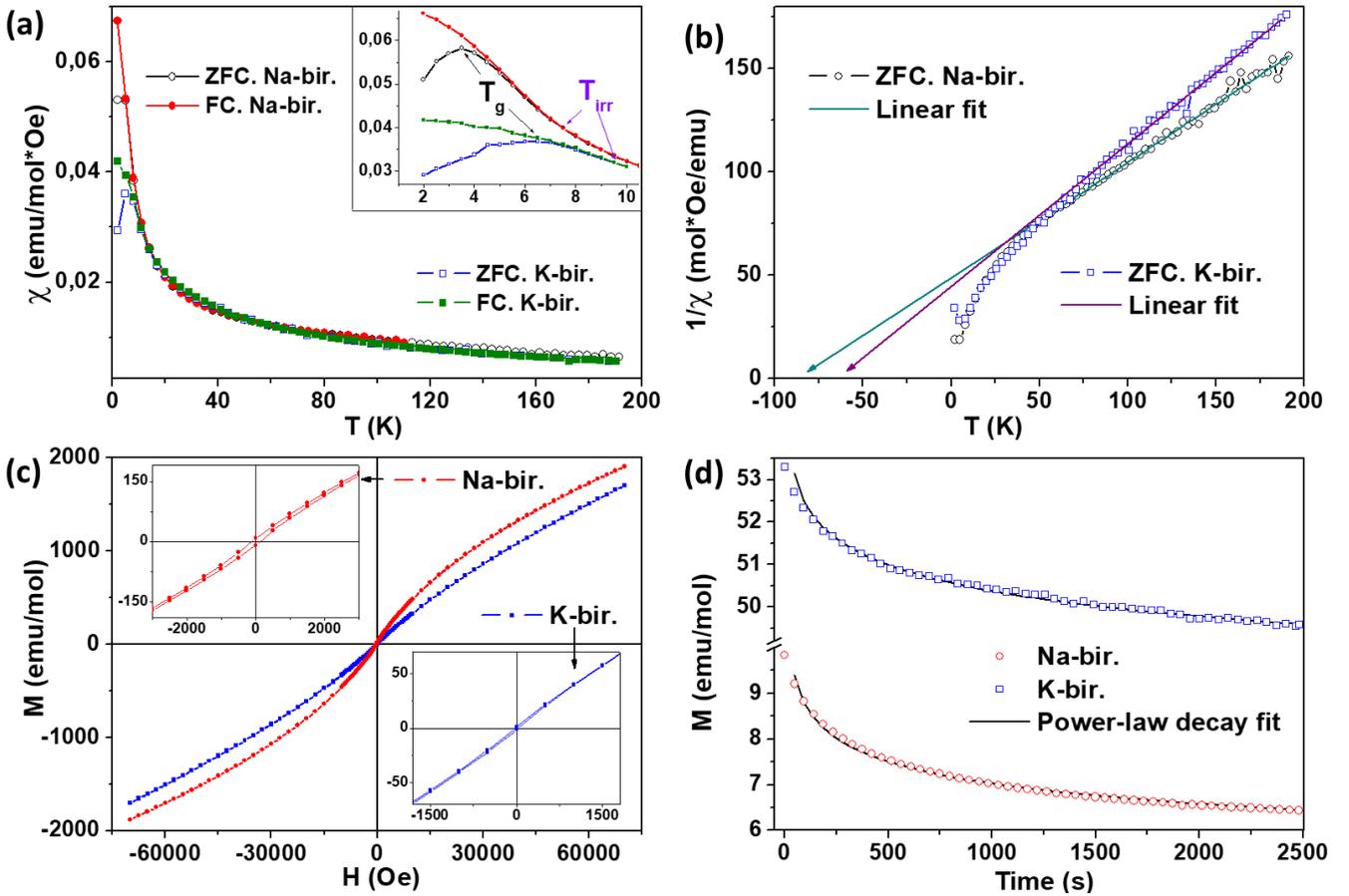

FIG. 3. (a) Temperature dependence of ZFC (open symbols) and FC (solid symbols) DC magnetic susceptibility of Na-bir (black and red) and K-bir (blue and green) measured on warming in a field of 500 Oe. The inset shows the DC susceptibility of samples in the 2-10 K range. (b) ZFC inverse DC susceptibility of Na-bir (black open symbols) and K-bir (blue open symbols) as a function of temperature at 500 Oe. The lines are linear fits to the experimental data above 50 K using the Curie-Weiss law. (c) Magnetization vs applied DC field measurement at 2.5 K for Na-bir (red) and at 5 K for K-bir (blue). The insets show closer views of the low-field region. (d) Magnetization decay measured by cooling Na-bir (red) and K-bir (blue) under a 1 T field to 2.5 K for Na-bir and 5 K for K-bir, then removing the field and measuring the magnetization as a function of time. The curves are fits to the experimental data using the power-law decay formula (Eq. (2)).

## C. AC susceptibility

To further investigate the origin of the peaks observed in the ZFC curves (Fig. 3(a)), the temperature dependence of the AC susceptibility $\chi_{AC}$ was measured over the temperature range 2-12 K at six different frequencies: 1, 10, 50, 100, 500, 1000 Hz for Na-bir and 50, 100, 250, 500, 750, 1000 Hz for K-bir; measurements at frequencies of 1 and 10 Hz for K-bir were too noisy for reliable results to be extracted. AC susceptibility measurements[28,29] are an important tool in the characterization of phase transitions and magnetic dynamics such as spin glass freezing, antiferromagnetic-paramagnetic transitions, and various magnetic relaxation processes including the irreversible movement of domain walls, narrow hysteresis loops in ferromagnets, spin-lattice relaxation in paramagnets, and relaxation of superparamagnets.[30,31,32,33,34]

The obtained AC-data (Fig. 4) consist of real and imaginary parts. The real component $\chi'(T)$ is in-phase with the oscillating field and probes reversible magnetization processes.[30] The $\chi'(T)$ curves of both samples exhibit a maximum that both decreases in height and shifts to higher temperature with increasing frequency, in the case of Na-bir this shift is more pronounced. The temperature of this maximum, which we refer to as $T_f$, is in both cases between 0.5 K and 1.5 K higher (depending on frequency) than $T_g$, at which the peak in the ZFC DC susceptibility is observed. Moreover, $T_f$ approximately corresponds to the temperature the inflection point is observed in the imaginary part $\chi''(T)$.[25,31,35] The imaginary component represents losses due to irreversible magnetization processes, which can involve different relaxations and energy absorbed from the applied field.[30] The maximum in $\chi''(T)$ shifts to higher temperature with frequency in case of Na-bir, but there is little or no shift for K-bir. At temperatures above $T_f$, $\chi''(T)$ tends toward zero. Such behavior is characteristic for a SG transition.[36,37]

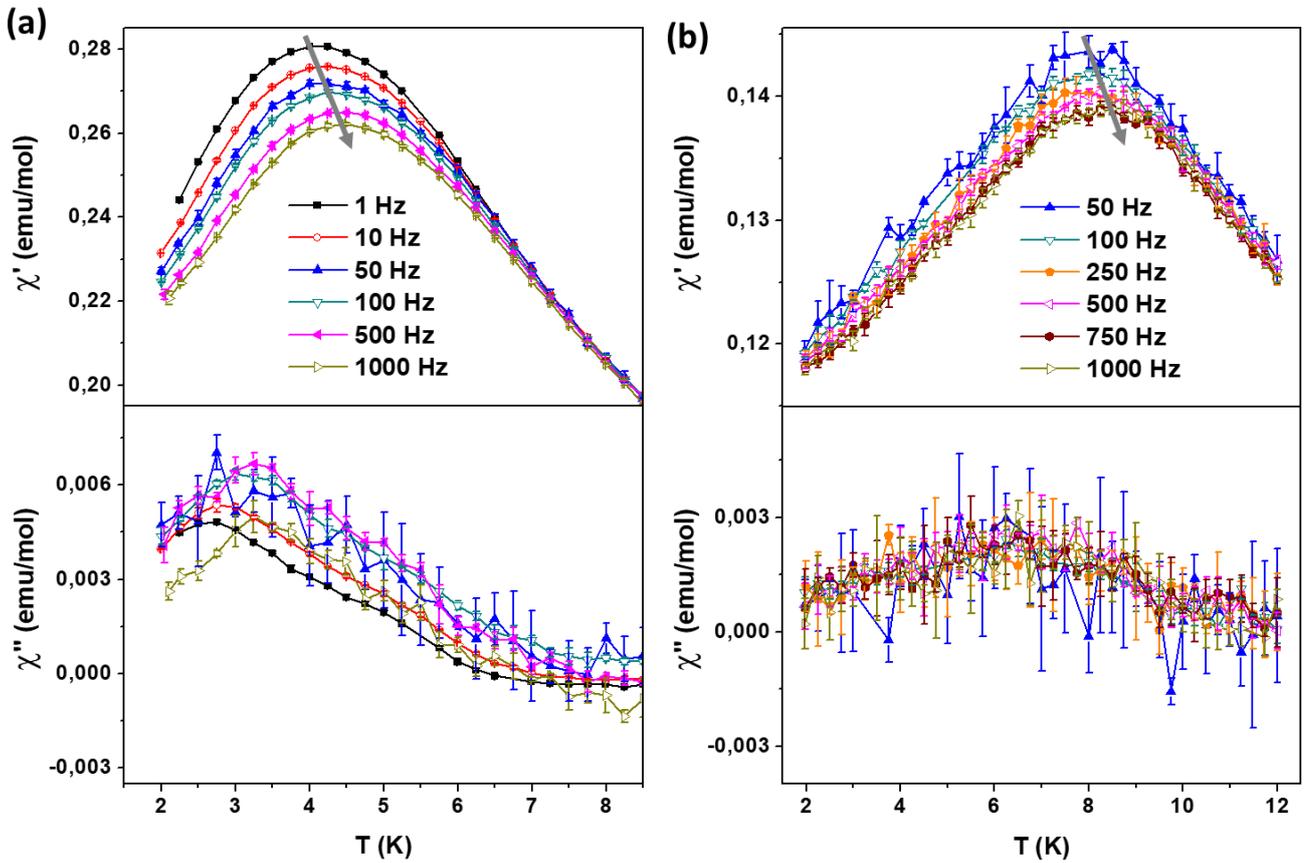

FIG. 4. Temperature dependence of the real and imaginary parts of the AC susceptibility at different frequencies, measured using a 3.8 Oe oscillating field and a 200 Oe DC field: (a) Na-birnessite at 1, 10, 50, 100, 500, 1000 Hz in the temperature range 2-9 K; (b) K-birnessite at 50, 100, 250, 500, 750, 1000 Hz in the temperature range 2-12 K.

A quantitative measure of the frequency dependence of the maximum in $\chi'(T)$ is given by the Mydosh parameter $\delta$:[34]

$$\delta = \frac{\Delta T_f}{T_f \times \Delta(Ln(\omega))} \quad (1)$$

Here $T_f$ is the freezing temperature, the frequency is $\omega = 2\pi f$, and $\Delta T_f$ is the difference between the maximum and minimum values of $T_f$. The Mydosh parameter allows magnetic states such as a SG[34] ($0.005 < \delta < 0.06$) and a non-interacting ideal superparamagnet ($\delta \leq 0.1$) to be distinguished.[38,39] The values of $\delta$ for Na/K-bir (Table II) correspond to the intermediate situation of a cluster glass (CG), also referred to as a reentrant spin glass, for which $\delta \sim 0.01$-

0.09.[25,26,40,41,42,43] This suggests that the maximum in $\chi'(T)$ is associated with randomly arranged, interacting magnetic clusters which become frozen below $T_f$. We note that a smaller value of $\delta = 0.007$ was obtained in the previous study of Na$_{0.36}$MnO$_2$·0.2H$_2$O by Bakaimi et al.,[14] corresponding to the canonical spin-glass regime.

### D. Magnetization versus applied field

The presence of spin clusters should be reflected in the shape of the magnetization (M) versus applied field (H) curve. The M-H curves measured below $T_f$ (at 2.5 K for Na-bir and 5 K for K-bir) exhibit an "S" shape in both cases (Fig. 3(c)), which taken alongside other evidence can also be a sign of spin glass/cluster glass systems in the frozen state, as was shown for other glassy compounds.[11,26,36,42] The magnetization does not reach saturation up to the highest applied field of 7 T (the expected saturation magnetization is $1.80 \times 10^4$ emu/mol for Na-bir and $2.01 \times 10^4$ emu/mol for K-bir). For Na-bir there is a narrow hysteresis loop but for K-bir any hysteresis is smaller than the step size in H (see insets in Fig. 3(c)). The existence of the hysteresis loop for Na-bir excludes a superparamagnetic ground state[44] and can be explained by the presence of competing FM and AFM exchange interactions in the glassy state.[11] The ferromagnetic interactions might arise from neighboring Mn$^{4+}$ cations, which in the triangular lattice of edge-sharing octahedra in the birnessite structure have Mn$^{4+}$-O-Mn$^{4+}$ bond angles of between 106° and 110° (Table IV), favouring FM superexchange.[45] Ferromagnetic double exchange is unlikely here due to the low-spin state of Mn$^{3+}$. Na-bir has a larger FM contribution, which is consistent with the higher average oxidation state of Mn$^{3.78+}$ compared to Mn$^{3.37+}$ for K-bir.

### E. Magnetization decay

To investigate the mechanism by which the system decays back to equilibrium after an external magnetic field is applied, magnetization decay measurements were carried out on both compounds (Fig. 3(d)). The measured data follow a power-law decay and can be fitted by the equation:[46]

$$M(t) = M_0 \times t^{-\alpha} \qquad (2)$$

Here $M_0$ is the maximum magnetization at the start of the measurement, and $\alpha$ is the decay parameter which is correlated with the decay rate. A higher decay parameter results in faster decay. This model can be applied to describe SG systems,[47] for which typical decay parameters[46] are of the order of $10^{-2}$. The fitted curves match the data well, with extracted parameters of $M_0 = 1.85 \times 10^{-5}$ emu/mol, $\alpha = 9.5 \times 10^{-2}$ for Na-bir, and $M_0 = 4.98 \times 10^{-5}$ emu/mol, $\alpha = 1.7 \times 10^{-2}$ for K-bir. Thus, the magnetization of K-bir decays significantly more slowly. These decay parameters imply relatively long time-scales, and are consistent with glassy behavior.[34]

### F. Dynamic scaling

To better understand the nature of the glassy phase of Na/K-bir, the dynamics of the SG state was studied by further analysis of the AC susceptibility measurements in Fig. 4. The frequency dependence of $\chi'(T)$ can be described by the critical slowing down formula[48] from dynamic scaling theory:

$$\tau = \tau_0 \times \left[\frac{T_f - T_g}{T_g}\right]^{-z\nu} \qquad (3)$$

Here $T_f$ was taken from the peak in $\chi'(T)$ for a given frequency $f$, and $T_g$ is the temperature at which the maximum in the ZFC DC susceptibility is observed, because $T_g$ can be regarded as the value of $T_f$ for infinitely slow cooling ($\lim_{f \to 0} T_f$).[26] The characteristic relaxation time of the dynamic fluctuations $\tau$ corresponds to the observation time $t_{obs} = 1/\omega = 1/(2\pi f)$ with the attempt frequency $\omega$, and the shortest time $\tau_0$ corresponding to the microscopic flipping time of the fluctuating entities. According to dynamic scaling theory, $\tau$ is related to the spin correlation length, $\tau \propto \xi^z$, and $\xi$ diverges with temperature as $\xi \propto [T_f/(T_f - T_g)]^\nu$ with the dynamic exponent $z$ and the critical exponent $\nu$.[49]

The left part of Fig. 5 shows a linear fit of $ln\tau$ vs $ln((T_f-T_g)/T_g)$, allowing values for $z\nu$ and $\tau_0$ to be obtained. These parameters are given in Table III. The $z\nu$ values for Na/K-bir are typical for glassy magnetism.[50,51,52] Typical values of $\tau_0$ for a canonical SG[52] lie in the range of ~ $10^{-12}$–$10^{-14}$ s; for cluster glasses[26,53,54] with slower dynamics, $\tau_0$ is in the range ~ $10^{-9}$–$10^{-11}$ s. The characteristic relaxation time of our compounds is of the order of ~ $10^{-12}$ for Na-bir and ~ $10^{-9}$ for K-bir. These parameters strongly suggest that both compounds exhibit a magnetic cluster glass state.

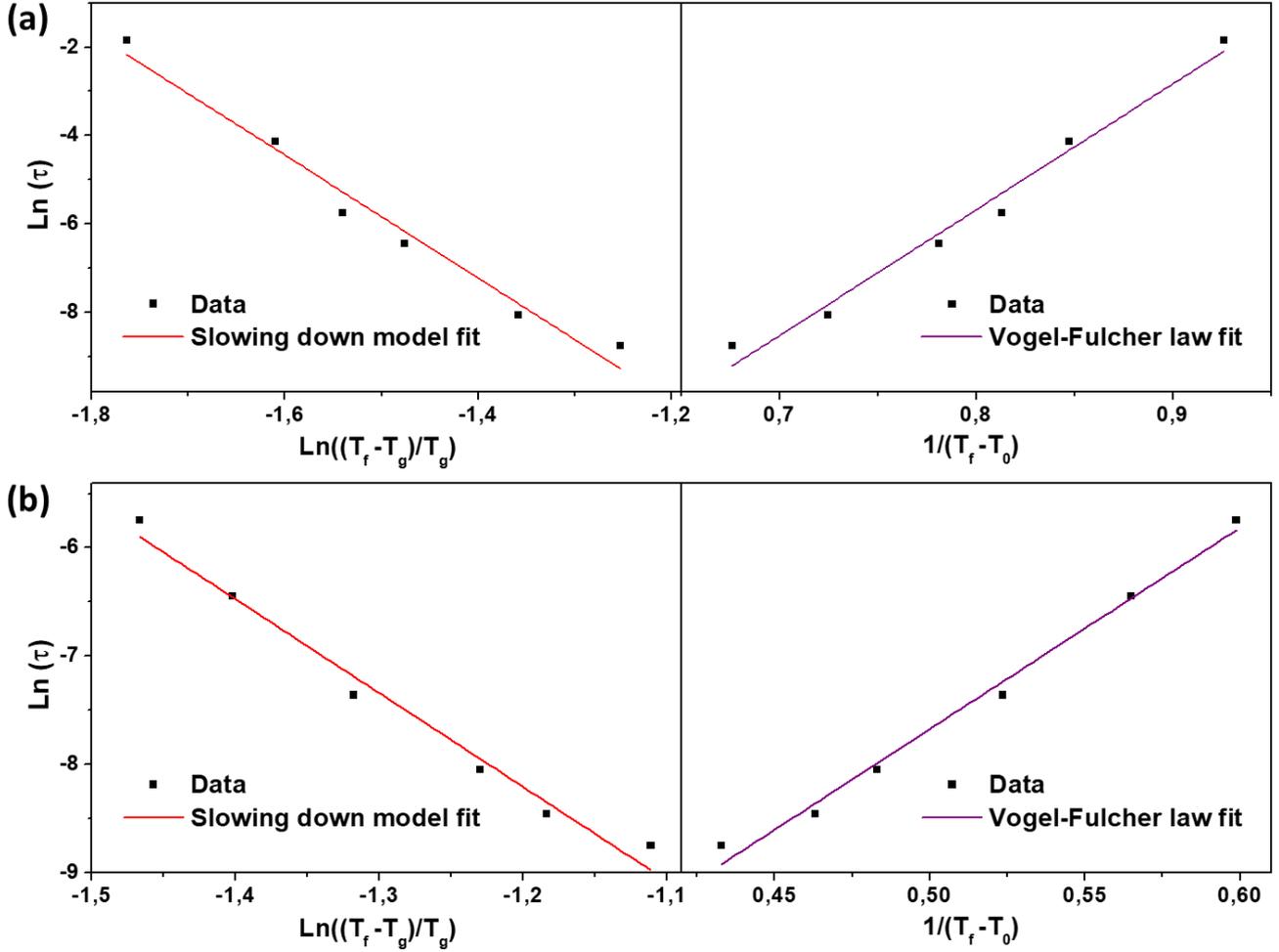

FIG. 5. Fits to AC susceptibility data (temperature at which peak in $\chi'(T)$ occurs for different frequencies) using the slowing down formula (Eq. (3)) and the Vogel-Fulcher law (Eq. (4)) for (a) Na-birnessite, (b) K-birnessite.

The dynamic magnetic properties of a glassy system can also be described by the Vogel-Fulcher law,[26,55] proposed for magnetically interacting clusters:

$$\tau = \tau^* \times exp\left[\frac{E_a}{k_B \times (T_f - T_0)}\right] \quad (4)$$

Here $T_0$ is a measure of the inter-cluster interaction strength, and $T_0$ is known as the Vogel-Fulcher temperature[34,36] and corresponds to the "ideal glass" temperature. Close to $T_0$, the Vogel-Fulcher law can be adjusted to match the power-law over a large frequency range:[52] $n\left[\frac{40\, k_B T_f}{E_a}\right] \sim \frac{25}{z\nu}$. This equation gives $E_a/k_B \sim 28$ K for Na-bir and $\sim 18.5$ K for K-bir. These values allowed the data to be fitted using Eq. (4) (right-hand panels of Fig. 5), yielding the parameters $\tau^*$, $T_0$ given in Table III. The extracted $\tau^*$ values lie in the range of $10^{-8}$–$10^{-13}$ s anticipated for glassy bulk systems with $Mn^{3+}/Mn^{4+}$ magnetic moments.[34,59] For K-bir, the smaller $\tau^*$ corresponds to a longer spin-flip time. Our values of $T_0$ are slightly lower than $T_g$ in the power-law model, as is the case for all SG systems.[52] A similar analysis was reported in the study of Bakaimi et al. on $Na_{0.36}MnO_2 \cdot 0.2H_2O$ [14] but the value of $\tau^*$ obtained was unexpectedly large, of the order of $10^{-6}$ s. The signal from the magnetic impurity phases present in the sample of Ref. 14 might have prevented reliable fitting. In other glassy systems containing mixed-valent $Mn^{3+}/Mn^{4+}$, values of $\tau^*$ ranging from $10^{-13}$ s to $10^{-9}$ s have been reported.[56,57,58] In Eq. (4), $T_0$ represents a measure of the coupling between the interacting entities,[59] where $T_0 \ll E_a/k_B$ corresponds to weak coupling and $T_0 \gg E_a/k_B$ to strong coupling. In the case of Na/K-bir, $T_0$ is much smaller than $E_a/k_B$ which implies the presence of weak interactions between the magnetic clusters.

TABLE III. Dynamic magnetic properties of Na/K-bir.

| Compound | Slowing down formula | | Vogel-Fulcher law | | |
|---|---|---|---|---|---|
| | $\tau_0$ (s) | $z\nu$ | $E_a/k_B$ (K) | $\tau^*$ (s) | $T_0$ (K) |
| $Na_{0.22}MnO_2 \cdot 0.39H_2O$ | $2.6 \times 10^{-12}$ | 13.9 | 28 | $4.6 \times 10^{-13}$ | 3.02 |
| $K_{0.6}MnO_2 \cdot 0.48H_2O$ | $8.4 \times 10^{-9}$ | 8.7 | 18.5 | $4.3 \times 10^{-8}$ | 6.33 |

## IV. CONCLUSIONS

In summary, we have synthesized the phase-pure birnessites $Na_{0.22}MnO_2 \cdot 0.39H_2O$ and $K_{0.6}MnO_2 \cdot 0.48H_2O$ by the sol-gel method and investigated their magnetic properties. DC magnetic susceptibility measurements show that antiferromagnetic interactions dominate. A bifurcation of the ZFC-FC magnetic susceptibility occurs at low temperatures with a distinct peak in the ZFC branch that suggests the presence of magnetic irreversibility. AC magnetic susceptibility and magnetization decay measurements demonstrate that the magnetic irreversibility likely originates from formation of a cluster glass state below the glass freezing temperature. The position of a peak in the real part of the AC susceptibility, accompanied by a peak in the imaginary component, is frequency-dependent and can be described by both the standard critical slowing-down formula and the Vogel-Fulcher law, which confirm the presence of a cluster glass state with weak interactions between clusters at low temperatures.

A comparison of our results with previously reported data on Na-birnessites with different compositions requires us to consider that a higher alkali cation content in the inter-layer space leads to a larger proportion of $Mn^{3+}$ cations. This results in an increased likelihood of Jahn-Teller distorted $MnO_6$ octahedra, as in the case α-$Na_{0.9}MnO_2$.[15,16] In the opposite situation where there is a large deficiency of alkali cations and a correspondingly larger proportion of $Mn^{4+}$ cations, a larger FM contribution is observed due to $Mn^{4+}$-O-$Mn^{4+}$ superexchange with a bond angle close to 90°. This is apparent from our magnetization versus applied field measurements where Na-bir exhibits a higher magnetization and wider hysteresis loop compared with K-bir. A larger proportion of $Mn^{4+}$ also suppresses the Jahn-Teller distortion and leads to a $Mn^{3+}$ low spin state, whereas $Mn^{3+}$ adopts the high spin state in α-$Na_{0.9}MnO_2$.[15,16]

Finally, it should be mentioned that the terms *spin glass* and *cluster glass* encompass a very broad range of specific magnetic states. In many cases, identification of a spin glass/cluster glass is only the beginning of the investigation. For alkali manganites with the birnessite structure, further study of the origin of the magnetic clusters is recommended. Due to the high flexibility in the alkali cation content in the interlayer space of birnessite, it would be interesting to study how to achieve control of the alkali cation occupation and the amount of crystal water, as well as to perform ion exchange of the interlayer species with different alkali cations or other inorganic/organic species. This will open new opportunities to tune the magnetic frustration inherent to birnessites and to create new magnetic states.

## ACKNOWLEDGMENTS

This work was supported by the European Union's Horizon 2020 research and innovation program under Marie Sklodowska-Curie Individual Fellowship, grant agreement no. 833550. We would like to thank Dr. Pavan Nukala for assistance with the SEM images and Ing. Jacob Baas for technical advice.

## APPENDIX A

Results of the thermal analysis of the samples are presented in Fig. 6. Three main endothermic effects and corresponding mass losses can be identified in the differential scanning calorimetry (DSC) and thermogravimetric analysis (TG) curves in the temperature range 30-500 °C while the samples were heated under a flow of argon gas. The lowest temperature feature corresponds to surface water evaporation. For Na-bir the mass loss is 5.2% and according to the time derivative of the TG curve (DTG), the loss of water is completed at 95 °C with the maximum of the DSC curve at 84 °C. For K-bir the mass loss is 2.2% and is completed at 79 °C, with the maximum of the DSC curve at 70 °C. The second endothermal effect (maxima in the DSC curve at 115 °C for Na-bir and 127 °C for K-bir) corresponds to the release of interlayer crystal water with mass losses of 7.1% for Na-bir and 7.2% for K-bir. A subsequent, much smaller mass loss of 2.3% for Na-bir and 0.5% for K-bir does not coincide with an obvious maximum in the DSC curve and might correspond to the release of remaining OH groups in the interlayer space. The last endothermal effect, shown by a broad minimum in the DTG curves and a small mass loss of 1.6% for both compounds (between 351 – 455 °C for Na-bir and 267 – 411 °C for K-bir) corresponds to a decomposition of the birnessite structure to $Mn_2O_3$ with the release of oxygen.

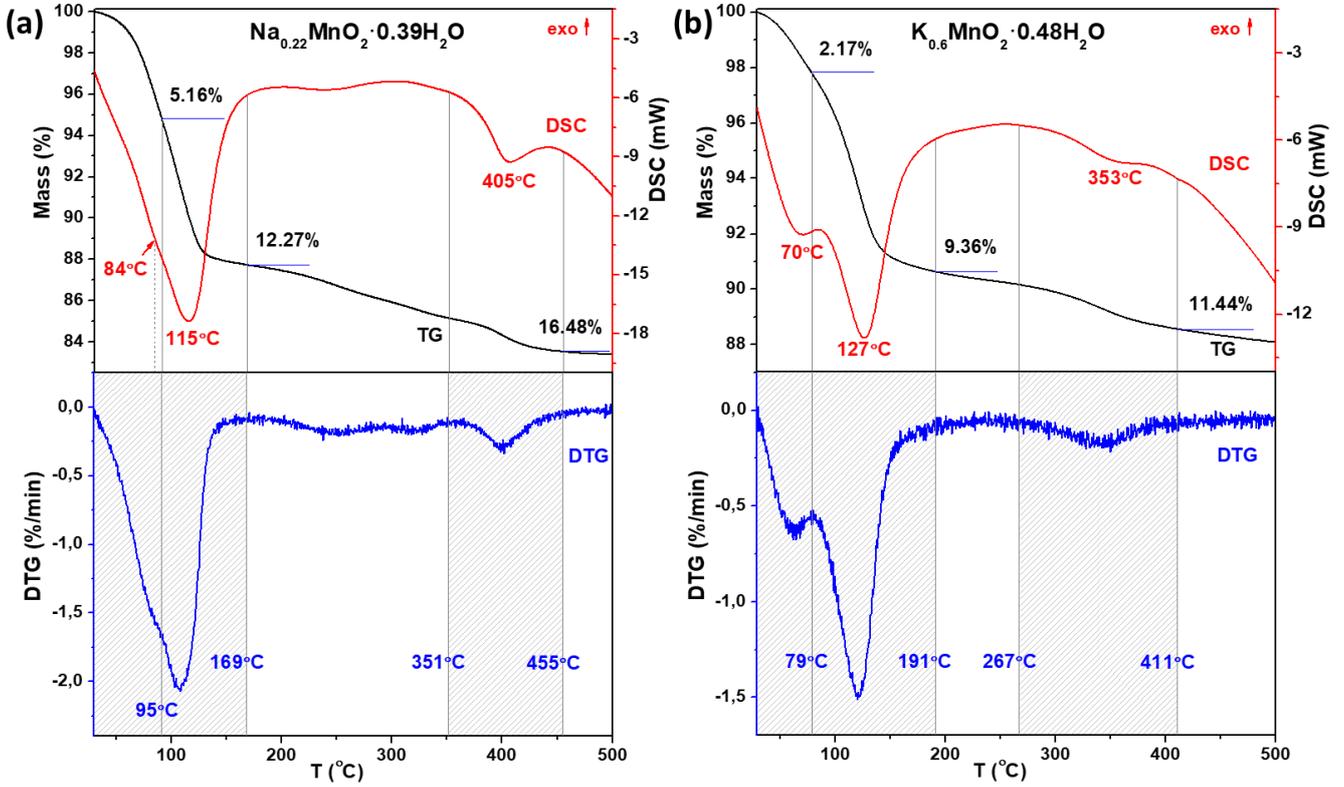

FIG. 6. Thermal analysis of (a) $Na_{0.22}MnO_2 \cdot 0.39H_2O$, (b) $K_{0.6}MnO_2 \cdot 0.48H_2O$. The TG and DSC data (top) and the DTG data (bottom) are represented by the black, red and blue lines respectively.

The oxidation state of the manganese atoms in the Na/K-bir samples was determined by a back-titration method.[60] Here 0.03 g of Na/K-birnessite was dissolved in 5 mL of 0.5 M aqueous sodium oxalate solution together with 10 mL of 1 M aqueous $H_2SO_4$ solution. The manganese ions are reduced to $Mn^{2+}$ and the oxalate ions are oxidized to produce $CO_2$ and $H_2O$:

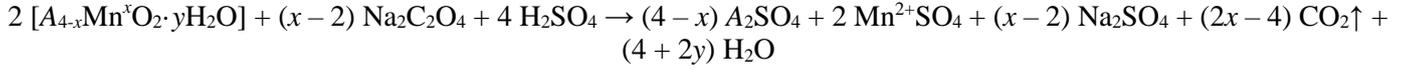

2 $[A_{4-x}Mn^xO_2 \cdot yH_2O]$ + (x – 2) $Na_2C_2O_4$ + 4 $H_2SO_4$ → (4 – x) $A_2SO_4$ + 2 $Mn^{2+}SO_4$ + (x – 2) $Na_2SO_4$ + (2x – 4) $CO_2\uparrow$ + (4 + 2y) $H_2O$

Here $A$ is $Na^+$ or $K^+$ cations in the birnessite compound, $x$ is the average oxidation state of the Mn ions, and $y$ is the amount of crystal water in the structure.

The unreacted oxalate was then back-titrated with 0.025 M aqueous $KMnO_4$ solution by the following reaction:

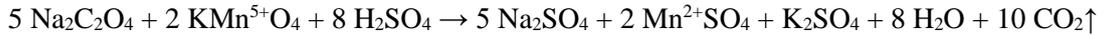

5 $Na_2C_2O_4$ + 2 $KMn^{5+}O_4$ + 8 $H_2SO_4$ → 5 $Na_2SO_4$ + 2 $Mn^{2+}SO_4$ + $K_2SO_4$ + 8 $H_2O$ + 10 $CO_2\uparrow$

Accordingly to the first chemical reaction the amount of sodium oxalate used to reduce the $Mn^{3+}/Mn^{4+}$ in the dissolved birnessite ($N_{1ox}$) corresponds to $N_{1ox} = \frac{x-2}{2} \times N_{bir}$, where $N_{bir}$ is the amount of used birnessite.

Or that equation can be shown as:

$$\frac{m_{bir}}{((4-x) \times M_A + M_{Mn} + 2 \times M_O + y \times M_{H_2O})} = \frac{2}{x-2} \times N_{1ox}$$

And the number of moles of unreacted oxalate can thus be determined based on the second reaction, which corresponds to 2.5 times the number of moles of permanganate used.

The values of $x$ and $y$ were obtained by combining the equations above and the equation below for the percentage mass loss of water during heating of the samples (see above):

$$\frac{y \times M_{H_2O}}{(4-x) \times M_A + M_{Mn} + 2 \times M_O + y \times M_{H_2O}} = wt\%_{H_2O}$$

The calculated amount of water is 0.39 molecules per formula unit in the case of Na-bir and 0.48 molecules per formula unit for K-bir. The average manganese oxidation state is +3.78 for Na-bir and +3.37 for K-bir assuming no oxygen or manganese vacancies. This results in a stoichiometry of $Na_{0.22}MnO_2 \cdot 0.39H_2O$ and $K_{0.6}MnO_2 \cdot 0.48H_2O$.

## APPENDIX B

Figure 7 shows Rietveld fits to the powder XRD data of Na/K-bir. The measured diffractograms show relatively broad peaks, which is common for samples prepared by sol-gel synthesis due to the nanoscale size of the particles formed. In the case of Na-bir, the sample consists of a single phase with the monoclinic structure (space group C2/m) previously reported by Post and Veblen.[61] The K-bir sample is also single-phase, with peaks that are much broader than for Na-bir. The peak profiles can best be modelled by introducing a triclinic distortion (space group C$\bar{1}$) as reported by Lopano et al.,[62] which gives clusters of reflections close together.

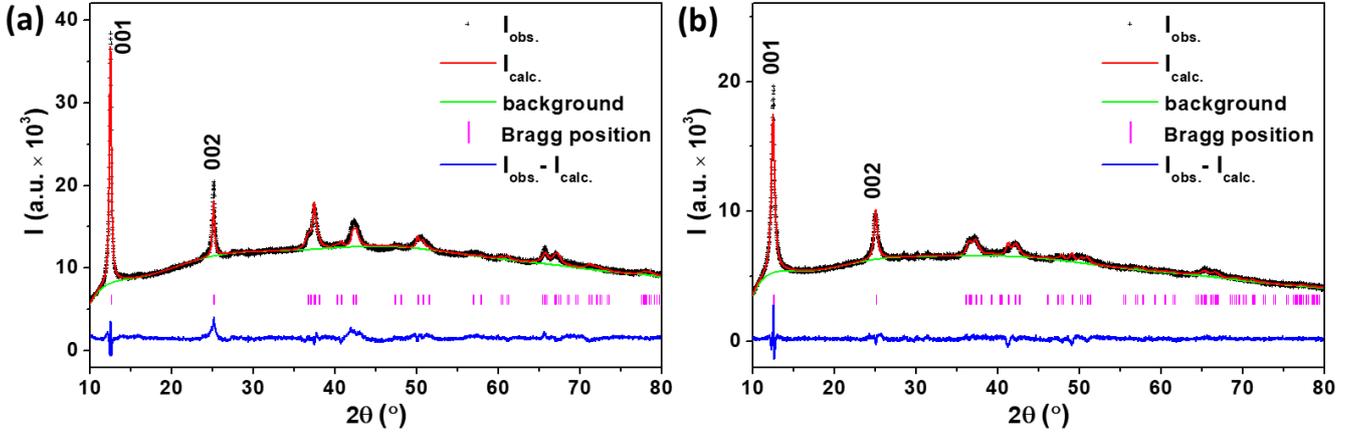

FIG. 7. Rietveld refinements using powder XRD data for (a) $Na_{0.22}MnO_2 \cdot 0.39H_2O$, (b) $K_{0.6}MnO_2 \cdot 0.48H_2O$. The observed data are denoted by black crosses, the calculated profile is the red line, and the difference profile is the lower blue line. The pink markers indicate the positions of allowed Bragg peaks.

The fitting was performed using the chemical compositions determined from the thermal analysis and back-titration method (Appendix A). The peak intensities could only be fitted well when a preferential orientation model was included in the fitting, accounting for a preferred packing of crystallites along the [001] direction. In both space groups the $Mn^{3+}$ and $Mn^{4+}$ cations occupy a single position at coordinates (0, 0, 0). Table IV lists the refined Mn-O bond lengths and the Mn-O-Mn bond angles. In the case of Na-bir the bond lengths are consistent with typical values for $Mn^{4+}$ cations, whereas the significantly longer bond lengths in the case of K-bir point to an intermediate valence state of $Mn^{3+}/Mn^{4+}$.

TABLE IV. Refined Mn-O bond lengths and Mn-O-Mn angles in Na/K-bir.

| Bond | Length (Å) | | Angle | Degree (°) | |
| --- | --- | --- | --- | --- | --- |
| | $Na_{0.22}MnO_2 \cdot 0.39H_2O$ | $K_{0.6}MnO_2 \cdot 0.48H_2O$ | | $Na_{0.22}MnO_2 \cdot 0.39H_2O$ | $K_{0.6}MnO_2 \cdot 0.48H_2O$ |
| Mn-O | 1.685(9) | 1.870(9) | Mn-O-Mn | 109.89(29) | 97.8(4) |
| Mn-O | 1.685(9) | 1.870(9) | Mn-O-Mn | 109.89(29) | 97.7(4) |
| Mn-O | 1.791(6) | 1.982(8) | Mn-O-Mn | 105.8(5) | |
| Mn-O | 1.791(6) | 1.942(9) | | | |
| Mn-O | 1.791(6) | 1.942(9) | | | |
| Mn-O | 1.791(6) | 1.982(8) | | | |

Figure 8 shows the structure of $K_{0.6}MnO_2 \cdot 0.48H_2O$, which is similar to the structure of $Na_{0.22}MnO_2 \cdot 0.39H_2O$ (Fig. 1). However, for K-bir it was not possible to unambiguously determine the distribution of $K^+$ cations and water molecules in the interlayer space. Therefore, the distribution of $K^+/H_2O$ was taken to be random in structural model used in the fitting.

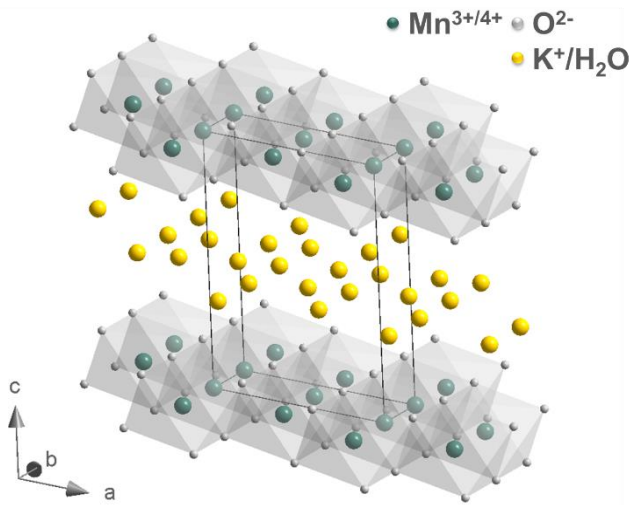

FIG. 8. Schematic representation of the crystal structure of $K_{0.6}MnO_2 \cdot 0.48H_2O$. The manganese and oxygen atoms are represented by dark green and grey spheres respectively; the positions of the potassium atoms and $H_2O$ molecules, sharing the same inter-layer site, are shown by yellow spheres. One unit cell is indicated by the dark grey lines.